\documentclass[aps,prl,twocolumn, superscriptaddress,showpacs]{revtex4}
\usepackage{graphicx}
\usepackage{psfrag}
\usepackage{subfigure}
\usepackage{verbatim}
\usepackage{color}
\usepackage{amsmath,amssymb}
\newcommand{\bea}{\begin{eqnarray}}
\newcommand{\eea}{\end{eqnarray}}
\begin{document}
\title{Toughness and damage tolerance of fractal hierarchical 
metamaterials}
\author{Ashivni Shekhawat}
\affiliation{Miller Institute for Basic Research in Science, University of California, Berkeley, 94720}
\date{\today}
\begin{abstract}
We present a theory for the toughness, damage tolerance, and tensile 
strength of a class of hierarchical, fractal, metamaterials.  We show 
that the even though the absolute toughness and damage tolerance 
decrease with increasing number of hierarchical scales, 
the specific toughness (toughness per-unit density) 
grows with increasing number of hierarchical 
scales in the material, while the specific tensile strength and 
damage tolerance remain constant. 
\end{abstract}

\maketitle
Three-dimensional (3D) metamaterials \textendash\ materials with 
designed structural features spanning several orders of magnitude in
length scales \textendash\ 
exhibit remarkable optical~\cite{soukoulis2011past, vignolini20123d, gansel2009gold},
thermal~\cite{PhysRevLett.112.055505,PhysRevLett.107.045901}, 
and mechanical~\cite{zheng2016multiscale, meza2014strong, 
jang2013fabrication,  2014vella, 2012Farr, kooistra2007hierarchical} properties. 
It has been demonstrated that such materials can have
large tensile strain ($\sim$ 20\%), elastic recovery, and 
can avoid the catastrophic failure mode usually associated with brittle
materials. This is particularly intriguing since the nanoscale
building blocks of these structures are often brittle. While 
considerable amount of work has been done towards understanding the 
modulus and compressive behavior of such materials~\cite{2014vella, 2012Farr, kooistra2007hierarchical}, the theoretical 
underpinnings of their fracture and flaw tolerance properties 
remain poorly understood. Further, toughness of such materials 
has never been investigated. In this Letter we present a theoretical 
framework for understanding the fracture toughness, tensile strength,
and flaw tolerance of a simple 
fuse network based model of such fractal, hierarchical materials. 
\par
Materials with structure at several length scales are, of course, 
not new. The biological world is full of such structural metamaterials~\cite{munch2008tough, Bonderer1069}, 
including bone, nacre, tooth, antler, wood etc. In fact, 
one would be hard pressed to find even one example of a biological material
which is monolithic in its structure across length scales. 
Most man-made materials are, by contrast, bland. For instance, metallic alloys
have no intrinsic structural features beyond the grain size, 
and most composites do not have features much below the 10-100$\mu$ range.
Further, even when traditional artificial materials
have structural features,
they are limited to one or two length scales, and lack the hierarchy of
length scales observed in their biological counterparts. However, 
recent technological developments in high-resolution large area 
additive manufacturing techniques have led to the development of 
samples with macroscopic dimensions approaching tens of centimeters, and a 
hierarchical structure spanning seven orders of magnitude with the finest 
structural features at the tens of nanometer scale~\cite{zheng2016multiscale}.
\par
Materials manufactured in this manner often have a 
fractal character~\cite{jang2013fabrication, meza2014strong} and 
do not admit a continuous stress field at all 
length scales. On the other hand, since 
continuum elasticity has no inherent length scales, continuous fields are 
bound to emerge 
at a length scale comparable to the largest length scale in the 
hierarchical structure. In other words, since averaging is possible 
at the largest length scale of the structure, 
thus continuum fields must emerge 
at this length scale. However, damage must initiate at the smallest 
length scale and propagate to the macroscopic scale upwards through the hierarchy.
Thus, it is important to understand how the continuum stress propagates 
via the hierarchy to the smallest length scales. We will develop 
a simple theory of this stress transfer and use it to develop scaling 
relations for toughness and strength of one simple model of a hierarchical 
metamaterial.
\par
\begin{figure}[tbp]
\begin{center}
\includegraphics[width=0.49\textwidth,angle=0]{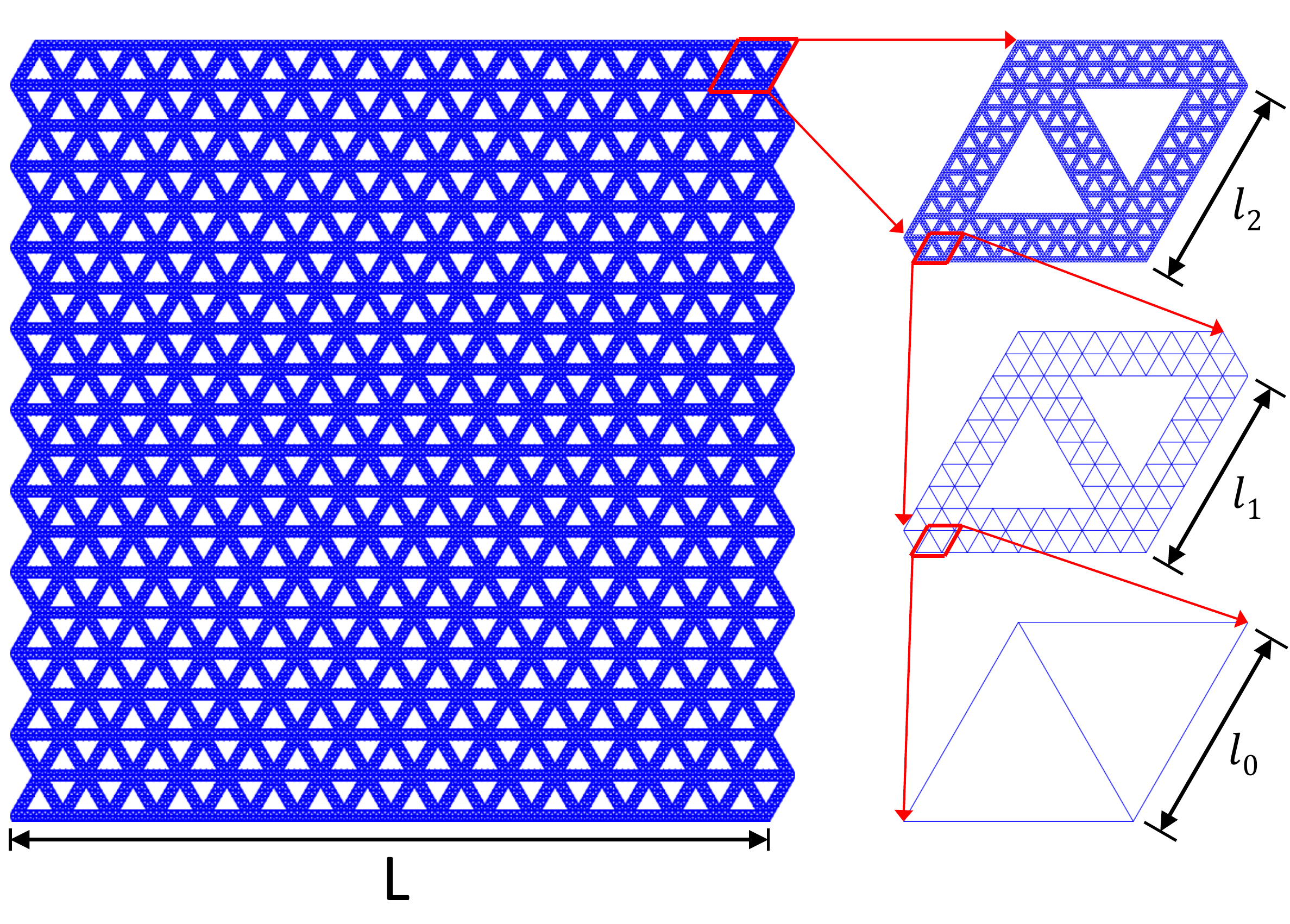}
\end{center}
\caption{{\bf A fuse network model of a hierarchical metamaterial.} 
At the longest length scale the material has a triangular lattice. However, 
each bond in the lattice has finer self-similar sub-structure. This 
refinement can continue ad infinitum; we restrict this figure to two 
levels and our later simulations to one level of refinement. }
\label{fig:1}
\end{figure}

Figure~\ref{fig:1} shows a schematic representation of our model material. 
The overall geometric dimension of the sample is $L$, while it has 
several hierarchical structural length scales, $l_i$ such that 
$l_0 \ll l_1 \ll \ldots \ll l_n \ll L$. Each length scale is related to 
the next scale by a magnification factor $m_i$ defined as $m_i = l_i/l_{i-1}$.
The fundamental structural unit of the model is the element at the 
smallest length scale $l_0$. This unit, and its interaction with 
the units connected to it, can be modeled in many ways. 
For instance, the units could be modeled as trusses, and the connections 
as pin joints, or the units could be modeled as beams, and the connections as 
rigid beam connectors. An even more realistic simulation would perhaps 
model the 
units as hollow thin-walled tubes. 
However, simulating such realistic models would be computationally 
taxing. Even a small two dimensional system
with $n=2$, $m=8$, and $L = 64\ l_2$ would
contain over $10^7$ units, thus making the simulation extremely challenging. 
We make several simplifications in order to enable meaningful simulations with modest computational resources. First, we study a two dimensional system. Second, we model the units as fuses, thus making a scalar approximation of 
tensorial elasticity~\cite{duxbury1987}. This approximation is known as 
the fuse network model. 
While this seems to be a drastic over-simplification, 
the fuse network model has been used widely to study brittle
fracture~\cite{arcangelis85, duxbury1986,
duxbury1987, beale1988, kahng88, mikko2006, shekhawat2012, shekhawat2013}. This model is formally equivalent to the anti-plane shear idealization of elasticity. 
Finally, we restrict our 
simulations to models with just one order of hierarchy, i.e., $n=1$. 

The basic fuse network model has been described in detail in several references~\cite{mikko2006, shekhawat2012}.
Briefly, in this model the current density is the analog of stress, the 
electric field that of strain, and the conductivity that of Young's modulus. 
Each fuse models a chunk of a brittle material that has a linear 
stress-strain (current density - electric field) characteristic up to a 
threshold stress, after which it breaks irreversibly. A fracture simulation 
is carried out by quasi-statically ramping the voltage across the 
network and removing fuses as their stress (current density) thresholds are 
reached. The voltage (displacements) at the nodes needs to be recalculated 
after each fuse rupture event. 
The network is said to be fractured when its conductivity drops 
to zero.
\par
We begin by considering the transfer of stress across the hierarchical 
length scales. In a network with $n$ hierarchical scales, 
the continuum stress, $\sigma_c$, emerges at a length scale 
comparable to the largest hierarchical scale $l_n$. If we imagine the 
next lowest length scale, $l_{n-1}$, as being monolithic (i.e., devoid of any 
further internal structure), then the stress at this scale can be related 
to the continuum stress at scale $l_n$ by a simple argument. The ratio 
of the area covered by the units at length scale $l_{n-1}$ and $l_n$ 
is given by $a_{n-1}/a_n = (6m_n - 9)/m_n^2$, where $m_n$ is the magnification
factor introduced earlier. Thus, we get 
$\sigma_c = \sigma_h^n = \sigma_h^{n-1}(6m_n - 9)/m_n^2$, where $\sigma_h^i$ 
is the effective hierarchical stress at the level $i$.
A similar idea for 
stress transfer via area-ratios is used in 
damage mechanics~\cite{lemaitre1984use, chaboche1988continuum}, 
however, in damage mechanics one does not usually consider a hierarchy 
of length scales. Continuing this argument,
we can obtain the relation between the continuum stress and the hierarchical 
stress at any level as 
\begin{equation}
\sigma_h^i = \sigma_c \prod_{k=1}^{n-i} \frac{m_{i+k}^2}{6m_{i+k}-9}.
\label{eq:stressTrans}
\end{equation}
The exact from of the above relation depends on the geometrical details, 
but the asymptotic relation $\sigma_h^i \sim \sigma_c m_{i+1}\ldots m_n$ is 
generic. If the magnification factors are constant, then in a 
structure with $n$ levels of hierarchy, the stress at the lowest 
level is given by $\sigma_h^0 \sim \sigma_c m^n$. 

\begin{figure}[tbp]
\begin{center}
\includegraphics[width=0.49\textwidth,angle=0]{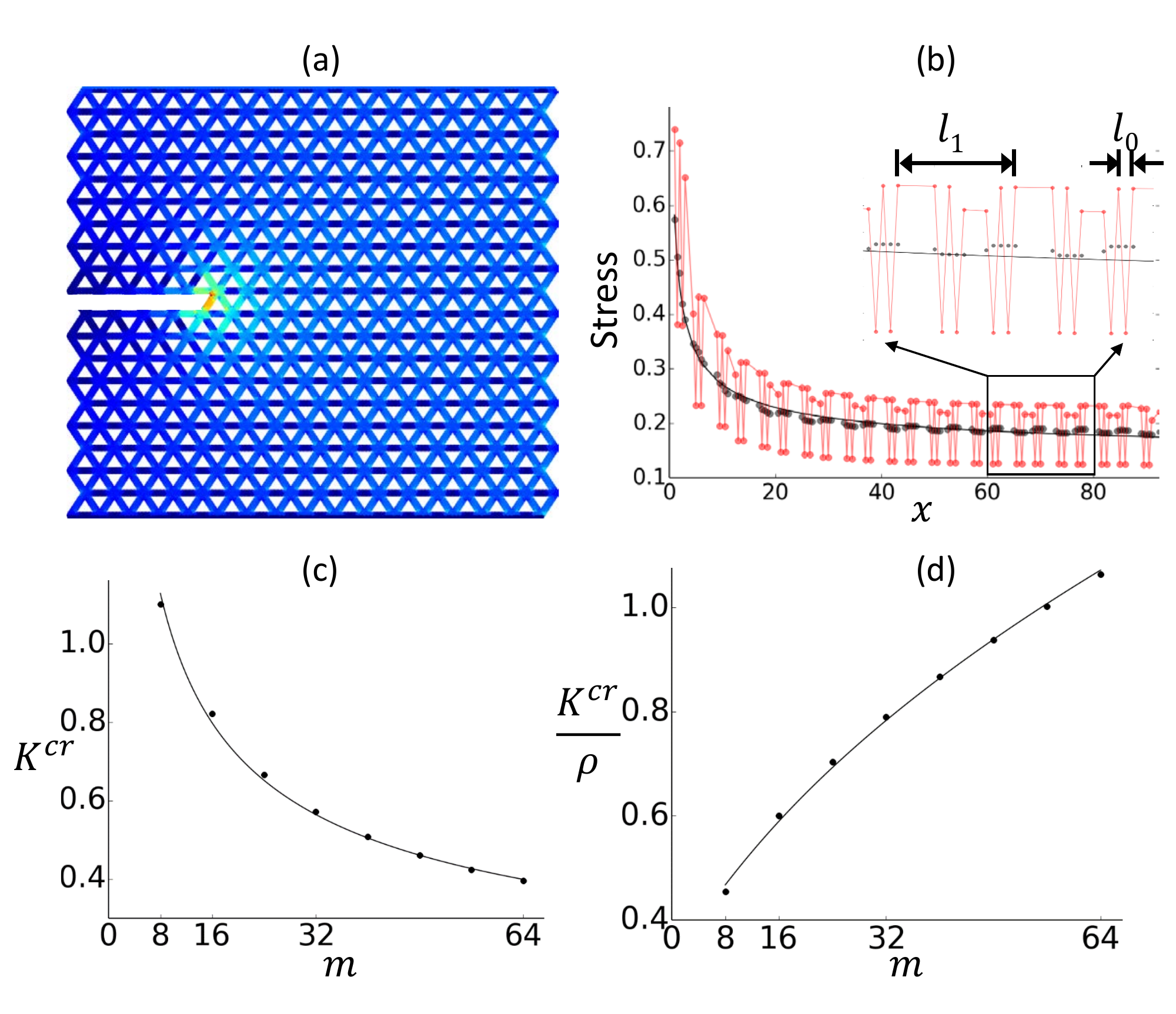}
\end{center}
\caption{{\bf Scaling properties of fracture toughness of metamaterials.}
{\bf (a)} 
A fuse network with one two structural length scales $l_0 \ll l_1 \ll L$.
{\bf (b)} 
Stress (current density) ahead of the crack tip 
as a function of distance from the crack tip.
The red circles show the current in the individual fuses. 
The zoomed inset identifies the periodic fluctuations at the 
hierarchical length scales. The 
black circles show the average current in one `truss-like' member 
at scale $l_1$ (note that this is not the continuum stress 
$\sigma_c$, rather it is an averaged hierarchical stress
$\langle \sigma_h^0 \rangle$). The black line is a fit to the 
form $\langle \sigma_h^0(x)\rangle \sim K/\sqrt{x} + T$. 
{\bf (c), (d)} Toughness (critical stress intensity factor), 
$K^{cr}$, and the 
specific toughness $K^{cr}/\rho$ as a function
of the magnification factor $m$. The simulation data is show by 
the circles while the curves are fitted to the asymptotic predictions
$K^{cr} = c_0\sigma^* l_0^{1/2}/\sqrt{m}$,
$K^{cr}/\rho = \sigma^* (l_0^{3/2}/\lambda) (c_1\sqrt{m} + c_2/\sqrt{m})$ 
where $c_0 = 3.19,\ c_1 = 0.13,\ c_2 = 0.29$ are fitted parameters.}
\label{fig:2}
\end{figure}

Consider now a system with a notch as shown in 
figure~\ref{fig:2}a. In order to propagate, the crack tip must cause 
damage at  
a distance $l_n$, where $l_n$ is the largest hierarchical length scale. 
The continuum stress a distance $l_n$ away from a crack tip is given by 
$\sigma_c \sim \frac{K}{\sqrt{l_n}}f(\theta) + T$, where 
$K$ is the stress intensity factor, 
$T$ is the first correction to the asymptotic stress 
(also known as the T-stress), and $f(\theta)$ is a function of the 
orientation. Ignoring the T-stress at short distances, the corresponding 
hierarchical stress at the smallest 
length scale is then given by  
$\sigma_h^0 \sim K m_1\ldots m_n/\sqrt{l_n}$ (Eq.~\ref{eq:stressTrans}). 
Fracture propagates when the 
stress at the lowest length scale reaches a critical value $\sigma^*$.
This critical value is the strength of the smallest 
building blocks of the structure. Using this, and the fact the largest
hierarchical length scale is given by $l_n = l_0 m_1\ldots m_n$, we 
get the scaling of the critical stress intensity factor as 
\begin{equation}
K^{cr} \sim \sigma^*l_0^{1/2}\prod_{k=1}^{n}\frac{6m_k-9}{m_k^{3/2}}.
\label{eq:Toughness}
\end{equation}
As before, the exact details above depend on the geometry of the structure, 
but the asymptotic relation $K^{cr} \sim \sigma^*(l_0/m_1\ldots m_n)^{1/2}$
is generic, and in particular if the magnification factors are constant,
then one obtains $K^{cr} \sim \sigma^*(l_0/m^n)^{1/2}$. Thus, the 
toughness of the material degrades as more hierarchical are added ($n$ is
increased) or as the magnification factor is made larger ($m$ is increased). 
While this sounds discouraging, we must keep in mind that both of these 
changes lead to a reduction in the density of the material. Thus, perhaps
a more meaningful property is the toughness per-unit density, or $K^{cr}/\rho$,
where $\rho$ is the density of the material. It can be shown that the 
density of our model hierarchical material is given by
\begin{equation}
\rho = \frac{2\sqrt{3}\lambda}{l_0}\prod_{k=1}^{n}\frac{6m_k-9}{m_k^2},
\end{equation}
where $\lambda$ is the linear mass density of the elements at scale $l_0$.
Thus, the toughness per-unit density is expected to scale as 
$K^{cr}/\rho \sim \sigma^*\frac{l_0^{3/2}}{\lambda}(m_1\ldots m_n)^{1/2}$.
Thus while the toughness of material decreases with 
increasing number of hierarchical 
levels and magnification factors, the specific toughness 
shows the opposite trend. 
\par
We test these predictions with numerical simulations of the fuse
network model with $n=1$ and $m=8,\ 16,\ 24,\ldots, 64$ 
as shown in figure~\ref{fig:2}a.
Thus, the simulated model has two internal length scales $l_0$, 
and $l_1$, with $l_0 \ll l_1 (=ml_0)$.
The units of length, mass, and stress are arbitrary 
in these simulations, and we choose them such that the elemental stress
threshold, linear mass density, and  
length are set to one, i.e., $\sigma^*=1,\ \lambda = 1$, and $l_0$ = 1.  
Figure~\ref{fig:2}b confirms the claim 
that the asymptotic form of the 
stress field in the hierarchical fuse network follows the 
predictions of linear elastic fracture mechanics, i.e., 
$\sigma(x) \sim K/\sqrt{x} + T$ where $x$ is the distance from the crack tip.
Figures~\ref{fig:2}c, d confirm the predicted scaling of the toughness 
(equation~\ref{eq:Toughness}) and the specific toughness with 
the magnification factor $m$. 
The critical stress intensity factor for the simulation is evaluated by fitting the continuum stress ahead
of the crack at each step in simulation to the form $\sigma_c(x) = K/\sqrt{x} + T$, and defining the maximum value of the 
fitted $K$ to 
be the critical stress intensity factor $K^{cr}$.
\par
We next focus on the damage tolerance and strength of the material. 
We consider a system that has some pre-existing damage. This is to say 
that any fuse at the smallest length scale $l_0$ can be missing with 
a small probability $p$. This random damage introduces a measure 
of disorder in the system. The strength 
of a given realization of this disordered structure is defined as the 
maximum continuum current density that it can support before fracturing and 
becoming non-conductive. Our objective is to establish a relationship 
between the mean strength and parameters such as the damage threshold $p$, 
number of hierarchical levels $n$, and the magnification factors $m_i$. 
We restrict ourselves to small values of $p$. The variation 
of mean strength with $p$ has the interpretation of quantifying 
the tolerance of the structure to microscopic damage and flaws. 
\par
\begin{figure}[tbp]
\begin{center}
\includegraphics[width=0.49\textwidth,angle=0]{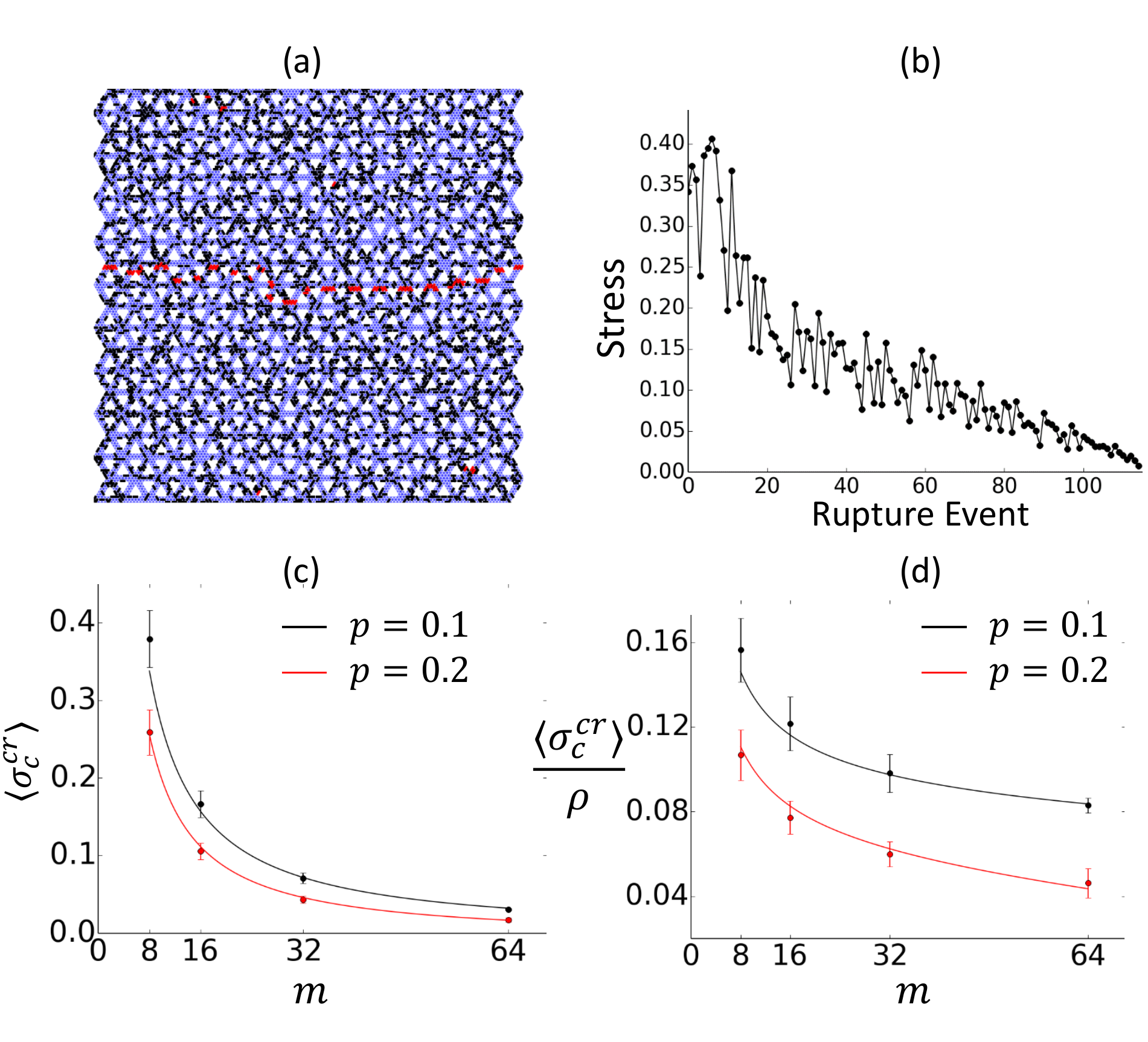}
\end{center}
\caption{{\bf Scaling properties of strength of disordered hierarchical 
metamaterials.} (a) A snapshot of disordered, fractured sample with $p=0.1$.
The pre-existing damage 
(fuses removed with probability $p$ prior to loading) 
is show in black. The fuses that break upon loading the system are 
shown in red. The final crack in the system is clearly visible. 
{\bf (b)} The evolution of stress in the system shown in (a); the abscissa is
the number of fuses broken till that point in the simulation. The system
shows a small amount of damage tolerance as the stress needed to break new 
fuses initially increase before peaking at about 0.40. 
{\bf (c), (d)} Mean strength and the 
mean specific strength as a function of the magnification $m$ at 
two levels of damage, $p = 0.1,\ 0.2$. 
The data is shown by the circles while 
the curves are a joint fit to the predictions  
$\langle \sigma_c^{cr} \rangle \sim (c_1/m)( -c_2\log p - \log m)^{1/2}$,
$\langle \sigma_c^{cr} \rangle/\rho \sim (c_3 + c_4/m)( -c_2\log p - \log m)^{1/2}$, with fitted parameters $c_1=1.21,\ c_2 = 3.08,\ c_3 = 0.05,\ c_4=0.15$.}
\label{fig:3}
\end{figure}

The effect of microscopic damage on the strength of 
fuse networks (and brittle materials in general)
is a well studied problem~\cite{duxbury1987, duxbury1986, shekhawat2012, harlow1978, phoenix1997}. However, 
the structures studied previously were monolithic.  
By contrast, the fractal nature of 
our model material gives rise to new phenomenology.
Figure~\ref{fig:3}a shows a snapshot of a simulation of fracture in 
a network with one order of hierarchy ($n=1,\ m=8$), and figure~\ref{fig:3}b
shows the corresponding evolution of stress. Monolithic (non-hierarchical)
fuse networks are known to be extremely brittle 
with a monotonically 
decreasing stress curve. However, due to the hierarchical nature, 
the stress curve shown in figure~\ref{fig:3}b is not monotonically 
decreasing. A non-monotonic stress curve suggests
that the material can 
sustain some damage before failure, 
and does not fail catastrophically. This is due 
to the fact that the effective continuum stress decreases as one 
moves up the levels of hierarchy, 
thus progressively higher stress is needed to propagate the 
damage through the material. 
\par
Like any brittle material, the hierarchical fuse network fractures via 
crack nucleation at a rare large crack-like flaw. 
This `rare large crack-like flaw' is in turn 
created by statistical fluctuations that lead 
to several neighboring 
fuses being damaged. If such a crack-like flaw has a length $l_c$, then
the continuum stress at which it starts propagating is given by 
$\sigma_c^{cr} \sim K^{cr}/\sqrt{l_c}$. Since we have demonstrated that the 
specific toughness, $K^{cr}$, decays as $(m_1\ldots m_n)^{-1/2}$, 
one might assume that the mean strength will follow a similar trend. 
However, we will show that this intuition is incorrect, 
and the mean strength decays much 
faster as $(m_1\ldots m_n)^{-1}$. As we shall show, 
this is due to the fact that 
as the number of hierarchical levels or the value of the magnification
factor grows, it takes lesser and lesser amount of damage to produce 
cracks of the same fixed length. 
\par
We begin by estimating the probability of having a crack of the 
length $l_c = r l_n$ in the material. We estimate this probability in a 
bottom up manner. At the shortest length scale the probability of
creating a crack of length $l_0$ is simply given by $p$. One scale
up, one has to remove at least 5 fuses to create a crack of length 
$l_1$, however, this can be done at any of $m_1$ positions. Thus,
the probability of creating a crack of length $l_1$ is of the order 
of $p^5 m_1$. Another scale up, 5 cracks of length $l_1$ need to be 
created in order to create a crack of length $l_2$, and this can be done
at any one of $m_2$ lattice sites. Thus, the probability of creating 
a crack of length $l_2$ scales as $(p^5 m_1)^5 m_2$. Proceeding in this 
manner, the probability of creating a crack of length $l_n$, at the 
longest hierarchical scale is of the order of 
$P_c(l_n) \sim p^{5^n}m_1^{5^{n-1}}m_2^{5^{n-2}}\ldots m_n$. Thus, 
the probability of creating a crack of length $r l_n$ scales 
as $P_c(r l_n) \sim P_c(l_n)^r$.
\par
We next estimate the mean length of the longest crack in the 
system~\cite{duxbury1986,shekhawat2012}. 
Let the length of such a crack be $r l_n$. 
Since such a crack appears once per lattice, and the lattice has 
$(L/l_n)^2$ nucleation spots, thus we get $(L/l_n)^2 P_c(r l_n) \sim 1$.
This is to say that the expected number of such cracks in the system 
is 1. Upon simplification, we find the following expression for the 
mean length of the longest crack in the lattice
\begin{equation}
\langle l_c \rangle \sim \frac{-2l_n\log(L/l_n)}{5^n\log p + 5^{n-1}\log m_1
 + \ldots + \log m_n}.
 \label{eq:crackLen}
\end{equation}
Combining equations~\ref{eq:Toughness},~\ref{eq:crackLen}, and using
$\langle \sigma_c^{cr} \rangle \sim K^{cr}/\sqrt{l_c}$, 
we get the following expression for the scaling of the mean 
strength with the various parameters of interest
\begin{multline}
\langle \sigma_c^{cr} \rangle \sim \frac{\sigma^*}{m_1\ldots m_n}\\
 \times \left(\frac{-5^n\log p - 5^{n-1}\log m_1
  \ldots - \log m_n}{2\log(L/l_n)}\right)^{1/2}.
 \label{eq:meanStress}
\end{multline}

Specializing equation~\ref{eq:meanStress} for the case $n = 1$, 
and holding $L/l_n$ constant gives $\langle \sigma_c^{cr}\rangle \sim 
\sigma^*(-5\log p - \log m)^{1/2}/m$. We test this prediction 
by fitting the numerical observed mean strength to the functional
form $\langle \sigma_c^{cr}\rangle = 
c_1(-c_2\log p - \log m)^{1/2}/m$, where we introduce the fitting parameter 
$c_2$ to acknowledge the fact that equation~\ref{eq:meanStress} is 
an approximation, and the geometric factors like 5 are not expected to be
accurate. Statistical sampling is done by 
averaging over 100 realizations of the disorder at each value 
of $m$ and $p$. The fit shown in figure~\ref{fig:3}c 
yields a value $c_2 = 2.66$; one standard deviation error bars 
are indicated.  Proceeding similarly one can 
show that the specific strength scales as 
$\langle \sigma_c^{cr} \rangle/\rho \sim (c_3 + c_4/m)( -c_5\log p - \log m)^{1/2}$, and figure~\ref{fig:3}d shows a fit of this form to the numerical
data. Thus, the specific strength asymptotically approaches a constant, 
and has a very slow decay with the damage probability $p$. 


In summary, we have developed a theory for understanding the scaling 
properties of the strength, damage tolerance, and toughness of 
hierarchical, fractal, brittle metamaterials. Our theory is 
able to provide satisfactory explanation to the results of our 
simulations, and provides a principled way to relate the continuum 
scale properties, such as toughness and strength, to the hierarchical 
structure of the material. However, much work remains to be done in this 
nascent field. Our model ignores the bending and buckling of the 
elemental building blocks even though these modes of 
deformation can be important.
We have also ignored plastic deformation in the elements, which 
is another potentially important effect, particularly if the 
elements are metallic and have thickness $\gtrsim 100$nm.   
These limitations not withstanding, we hope that our results 
will lead to a better understanding of fracture in 
this fascinating class of materials. 
\bibliography{meta}
\end{document}